# Vanishing of the anomalous Hall effect and enhanced carrier mobility in the spin-gapless ferromagnetic $Mn_2CoGa_{1-x}Al_x$ alloys

Cheng Zhang[#], Shuang Pan[#], Peihao Wang[#], Yuchen Men, Xiang Li, Yuqing Bai, Li Tang, Feng Xu, Guizhou Xu[*]

*School of Materials Science and Engineering, Nanjing University of Science and Technology, Nanjing 210094, People's Republic of China*

**Abstract**

Spin gapless semiconductor (SGS) has attracted long attention since its theoretical prediction, while concrete experimental hints are still lack in the relevant Heusler alloys. Here in this work, by preparing the series alloys of $Mn_2CoGa_{1-x}Al_x$ (x=0, 0.25, 0.5, 0.75 and 1), we identified the vanishing of anomalous Hall effect in the ferromagnetic $Mn_2CoGa$ (or x=0.25) alloy in a wide temperature interval, accompanying with growing contribution from the ordinary Hall effect. As a result, comparatively low carrier density ($10^{20}$ cm$^{-3}$) and high carrier mobility (150 cm$^2$/Vs) are obtained in $Mn_2CoGa$ (or x=0.25) alloy in the temperature range of 10-200K. These also lead to a large dip in the related magnetoresistance at low fields. While in high Al content, despite the magnetization behavior is not altered significantly, the Hall resistivity is instead dominated by the anomalous one, just analogous to that widely reported in $Mn_2CoAl$. The distinct electrical transport behavior of x=0 and x=0.75 (or 1) is presently understood by their possible different scattering mechanism of the anomalous Hall effect due to the differences in atomic order and conductivity. Our work can expand the existing understanding of the SGS properties and offer a better SGS candidate with higher carrier mobility that can facilitate the application in the spin-injected related devices.

---

[*] Corresponding Author: E-mail: gzxu@njust.edu.cn

[#] These authors contribute equally to this work.

## I. INTRODUCTION

Since the proposal of spin gapless semiconductor (SGS) in concept[1], there is a wealth of *ab-initio* and experimental investigations on related Heusler alloys, such as $Mn_2CoAl$[2-6], CoFeMnSi[7,8], CoFeCrAl(Ga)[9-11], etc. Among them, $Mn_2CoAl$ alloy and its films have attracted most attention, giving its more ideal band structure and some electrical-transport hints of the spin gapless semiconducting properties [2,3,5]. But until now, the features account for a real SGS are still ambiguous, considering the observed large carrier density ($\sim 10^{20}$-$10^{22}$cm$^{-3}$ [3,12,13]), which actually approaches a metal, and low carrier mobility (<1 cm$^2$/Vs [3,5,13]), preventing its further application in spin-injection based devices. One of the key challenges is the precise control of the atomic order and microstructure in these ternary or quaternary intermetallic compounds.

For instance, in $Mn_2CoAl$, high resolution transmission electron microscope (HRTEM) analyses performed by Xu et al[13] have revealed significant non-stoichiometry and two-phases coexistence in the compound, which greatly hinders the realization of its claimed properties as SGS. Similar conclusions were made in the comprehensive structural characterization and analyses of its thin films[6]. Comparing to $Mn_2CoAl$, the iso-structural $Mn_2CoGa$ alloys receive much less attention concerning the SGS characteristics, possibly due to the predicted less perfect electronic structure[14]. However, as reported by Manna et al[15], unlike in $Mn_2CoAl$, vanishing anomalous Hall effect (AHE) was observed at low temperature in the single-crystalline $Mn_2CoGa$, which is a reflection of its trivial zero-band-gap feature across the Fermi level. Therefore, $Mn_2CoGa$ might be more promising to take advantage of its gapless feature in the application of efficient spintronic devices. Here in this work, we grow the series of polycrystalline alloys of $Mn_2CoGa_{1-x}Al_x$ (x=0, 0.25, 0.5, 0.75). In $Mn_2CoGa$ (x=0), the phenomenon of vanishing AHE is well reproduced and extensively studied, where substantially low concentration and high mobility of the charged carriers are identified. With growing Al content, the contribution from AHE gradually increases, until at x=0.75, it dominates the Hall signal like in $Mn_2CoAl$, leading to increased carrier density (by one order) and dramatically decreased mobility. The possible reason account for this distinct change caused by the mere substitution of neighboring main-group element is also analyzed.

## II. EXPERIMENTAL AND CALCULATION DETAILS

Polycrystalline samples of nominal composition $Mn_2CoGa_{1-x}Al_x$ (x=0, 0.25, 0.5, 0.75) were prepared by arc-melting of high-purity elemental metals in an argon atmosphere. The as-cast ingots were annealed in vacuumed quartz tubes at 1273 K for 5 days, followed by quenching into water. The crystal structure was identified by powder X-ray diffraction (XRD) with Cu-K$\alpha$ radiation (Bruker D8 Advance). Both the magnetic and electrical measurements were performed on the Physical Property Measurement System (PPMS, Quantum Design). For transport measurements, the samples were cut into regular rectangle shape with small thickness (0.3~0.5mm) and a four-probe method was applied.

In order to treat the substitution composition and antisite disorder, we apply the full-potential Korringa-Kohn-Rostoker method combined with the coherent potential approximation (KKR-CPA)[16,17], which is implemented in the AkaiKKR code [18]. In the calculation, the exchange correlation effect is treated with generalized gradient approximation (GGA) function.

## III. RESULTS AND DISCUSSION

Figure 1(a) exhibits the structure characterization of $Mn_2CoGa_{1-x}Al_x$ (x=0, 0.25, 0.5, 0.75, 1) alloys. The main XRD peaks of all the alloys can be well indexed to the Heusler structure of $Hg_2CuTi$ prototype (space group: $F\bar{4}3m$ (216)), but the superstructure lattice reflections of (111) and (200) are only discernable (around 30°) in $Mn_2CoAl$ and $Mn_2CoGa_{0.25}Al_{0.75}$ (only (200)) alloys, though rather weak due to the large background noise. However, this does not readily mean the lower atomic order in the Ga substituted alloys, since the superlattice reflections can also be affected by the individual atomic scattering. when the main group element is from the same period of the transition metal (here Mn, Co), it is difficult to ambiguously determine the correct ordered structure by mere x-ray diffraction[11,19]. For instance, superlattice reflections were absent in the XRD patterns of $Co_2FeZ$(Z=Ga, Ge) alloys, but the extended x-ray absorption fine-structure (EXAFS) technique revealed the existence of $L2_1$ high order[19]. The simulated XRD in Fig. 1(b) based on the fully-ordered structure further showed that the theoretical (111) and (200) peaks of $Mn_2CoGa$ are indeed much lower than that of $Mn_2CoAl$, almost undiscerned comparing to the main peak of (220). However, there were some cases that these reflections can be discerned in $Mn_2CoGa$, like reported in refs. [14, 20], probably due to improvements in sample quality or instrument conditions. The determined lattice constants of $Mn_2CoAl$ and $Mn_2CoGa$ are 5.83 and 5.86Å, respectively. Small expansion of the lattice in $Mn_2CoGa$ is caused by the larger atomic radio of Ga atom comparing to Al[14].

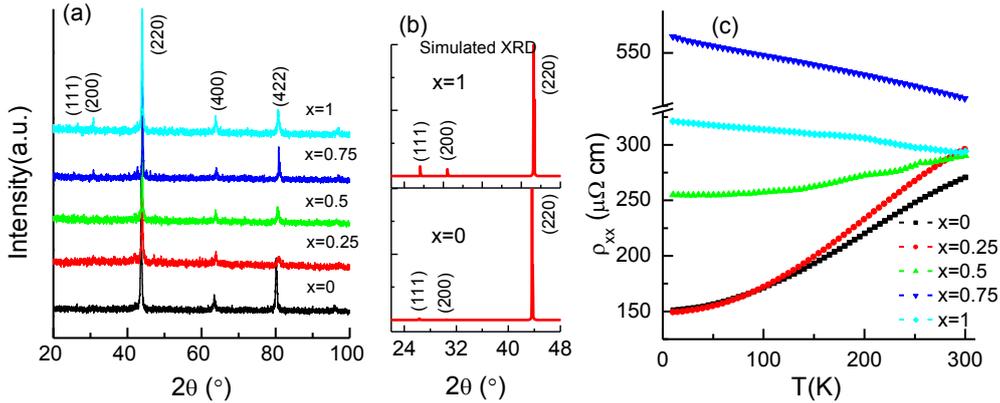

Fig.1 (a) XRD patterns of $Mn_2CoGa_{1-x}Al_x$ (x=0, 0.25, 0.5, 0.75, 1) alloys. (b) Simulated XRD pattern (only partly shown) based on the ideal full-order $Hg_2CuTi$ structure (structure details can be seen in Fig.6 and related text therein). The intensity of the main peak (220) is same for the two patterns. (c) Zero field resistivity ρ(T) curves of x=0-1.

The zero-field resistivity ρ(T) curves of $Mn_2CoGa_{1-x}Al_x$ (x=0, 0.25, 0.5, 0.75, 1) in Fig. 1(c) show distinct behavior with evolution of x. In x=0 and 0.25 with high Ga content, they behave like a typical metal, *i.e.*, with decreases of temperature, the resistivity continually drops, until at low temperature, it approaches to constant (residual resistivity). While with increasing Al content, the temperature coefficient of resistivity (TCR) starts to change from positive to negative. At x=0.75 and 1, the ρ(T) behavior is semiconducting-like, resembling the case generally reported in $Mn_2CoAl$[2,5,13]. However, it should be noted that the negative TCR is not a definitive characteristic of semiconducting. As the disorder scattering in high-resistivity transition metal alloys (>150 μΩ cm) could also result in this kind of behavior[21], which has been identified in quite a few intermetallic bulk and films[22-24]. As further proven in the work by H. *Pandey* et al[25], an enhancement in the atomic order of Heusler alloy thin film can lead to a transition of the TCR from negative to positive. Therefore, the evolution of ρ(T) behavior observed here can indicate the highly atomic order of Ga-rich alloy. On the other side, the overall higher resistivity in samples with larger Al proportion may possibly imply the increasing disorder in them.

In contrast to the different trend of zero-field resistivity, the magnetization behavior of the five $Mn_2CoGa_{1-x}Al_x$ samples varied little, as shown in Figs. 2(a-e). They all exhibited soft magnetization behavior with saturation fields at about 0.2T. The saturation magnetization ($M_S$) descends slowly when the temperature increased from 10K to 300K, owing to the high Curie temperature ($T_C$) of this series of alloys ($Mn_2CoAl$: Tc ~ 720K[2], $Mn_2CoGa$: Tc~ 740K[26, 27]). It is noticed that there exist subtle maximum at around 100K in the $M_S$ vs T curves, which may be caused by a small fraction of Mn clusters that are antiferromagnetic below ~100K in the α structure[28]. However, it is important to note that the change in $M_S$ is actually very small, limited to 5 emu/g. This indicates that the concentration of the possible Mn clusters is quite low and therefore, their influence on the overall properties of the matrix alloys is negligible. The experimental magnetic moment (at 10K) per formula unit is 2.12$\mu_B$ for $Mn_2CoGa$, deviates little to the theoretical value of 2 $\mu_B$. It further decreases with larger Al content, which may be caused by the growing anti-site disorder in the co-doped samples.

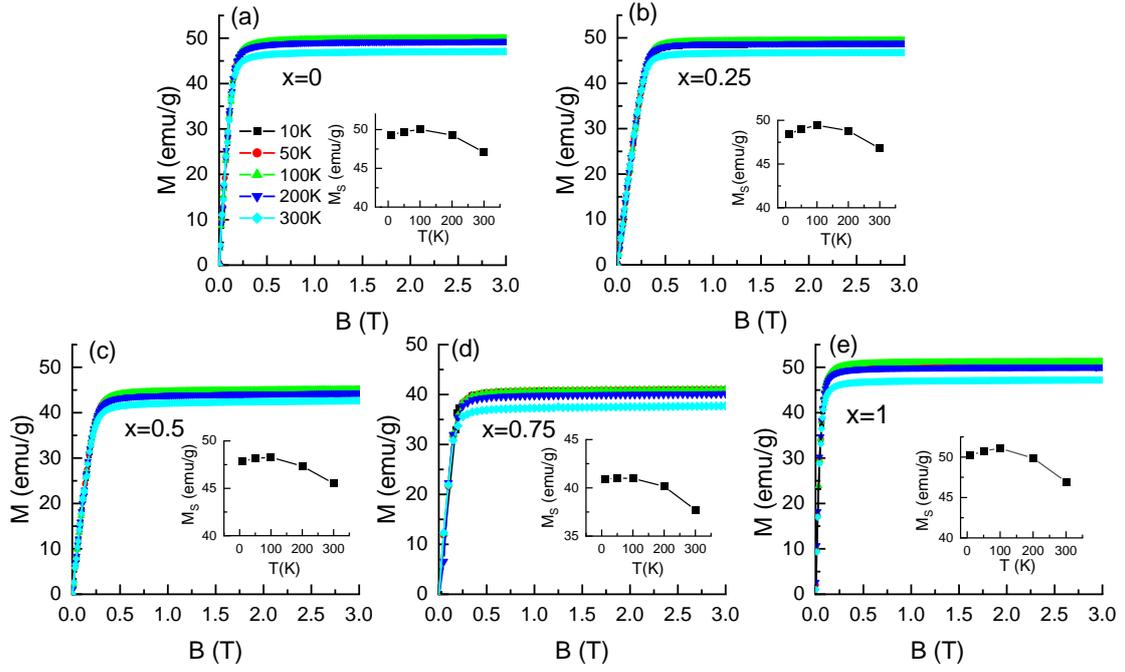

Fig. 2 (a-e) M-B curves for $Mn_2CoGa_{1-x}Al_x$ (x=0, 0.25, 0.5, 0.75, 1) alloys at various temperatures of 10-300K. Inset shows the temperature dependence of the extracted saturation magnetization.

In order to further investigate the transport properties of $Mn_2CoGa_{1-x}Al_x$ alloys, their field dependent Hall resistivities ($\rho_{xy}$) are measured for various temperatures ranging from 10K-350K. As shown in Fig. 3, distinct feature appears in the isothermal $\rho_{xy}(B)$ curves with evolution x. For x=0, *i.e.*, stoichiometric $Mn_2CoGa$, obvious AHE arises at 350 and 300K, which gradually vanishes with decreases of temperature, approaching to an almost linear line at 10K (inset). In the meantime, the slope of the high field $\rho_{xy}(B)$ curve increases, indicating the growing contribution from ordinary Hall effect (OHE) at low temperatures. This behavior is quite analogous to that reported in the single-crystalline $Mn_2CoGa$ alloy by Manna et al[15]. Here it is found that similar trend also occurs in x=0.25 and 0.5, only that the contribution of OHE becomes less evident with larger x. Until at x=0.75, the AHE dominates over entire temperature ranges, with negligible OHE contribution. The behavior of x=1 is similar to x=0.75, consistent with that observed in $Mn_2CoAl$ alloys[2,13].

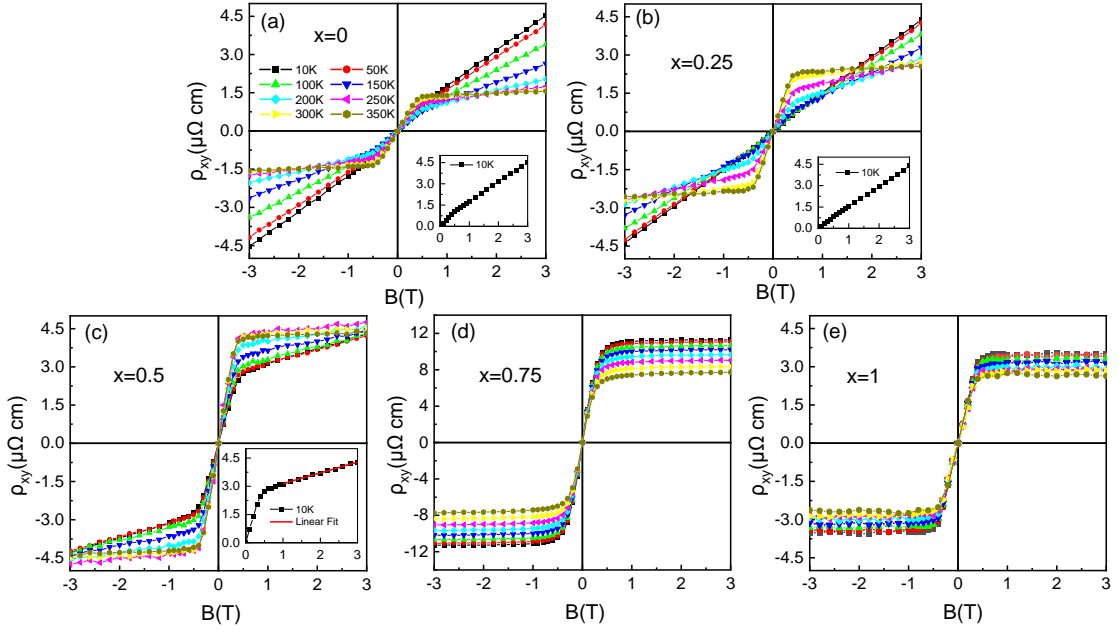

Fig. 3 (a-e) Magnetic field dependence of Hall resistivities ($\rho_{xy}$) for Mn$_2$CoGa$_{1-x}$Al$_x$ (x=0, 0.25, 0.5, 0.75, 1) alloys at various temperatures ranging from 10K-350K. Inset in (c) shows the high-field linear fit to calculate the related transport parameters (see the main text for details).

The anomalous Hall signal can be empirically expressed as $\rho_{xy}=\rho_{OHE}+\rho_{AHE}=R_0 B+R_s M$, where $R_0$, $R_s$ are the ordinary and anomalous Hall coefficient, respectively. As the magnetization $M$ approaches to a constant value ($M_S$) above the saturation fields, the anomalous Hall resistivity $\rho_{xy}^A$ and ordinary Hall coefficient $R_0$ can be obtained as the intercept and slope of the high-field linear fitting (see inset of Fig. 3c). The carrier concentration and mobility are further evaluated by $n = 1/eR_0$ ($e$ is the charge of the electron) and $\mu = \sigma_0/ne$ ($\sigma_0$ is the zero field longitudinal conductivity). The extracted temperature dependences of $\rho_{xy}^A$, $n$ and $\mu$ for all the samples are summarized in Figs. 4(a, c, d). It can be seen that in x=0 and 0.25, the anomalous Hall contribution is approximate zero below 200K, and grows finitely with increasing temperature. Correspondingly, their carrier concentrations are in a low scale (<10$^{20}$cm$^{-3}$) below 200K, but subsequently grows dramatically, nearly following a relation that obeyed in intrinsic semiconductors[29], that is $n \propto e^{-\Delta/2k_B T}$, where $\Delta$ represents the band gap of the material. For x=0.5, the carrier variation is similar to that of x=0 and 0.25, but with a higher base value. The fitted bandgap of them are all about 0.2 eV, approximating the nearly-zero bandgap of Mn$_2$Co(Ga, Al) alloys, as predicated in the literature[15,27] and following theoretical calculations. While for x=0.75 and 1, the carrier density is maintained at a high value (in the order of 10$^{21}$-10$^{22}$ cm$^{-3}$) and showed much gentle temperature dependence. Therefore, only for high Ga content, the carrier excitation mode behaves like a semiconductor. As the carrier mobility $\mu$ is inversely proportional to $n$, it presents a large magnitude at low temperature for x=0 and 0.25, which can reach ~150 cm$^2$/Vs, considerably higher than in regular metals. Comparatively, for x=0.5, since the contribution of AHE enhances, the $n$ increases and $\mu$ lowers. At x=0.75 and x=1, large $n$ and low carrier mobility are observed, in coincidence with that reported in $Mn_2CoAl$ [12, 13].

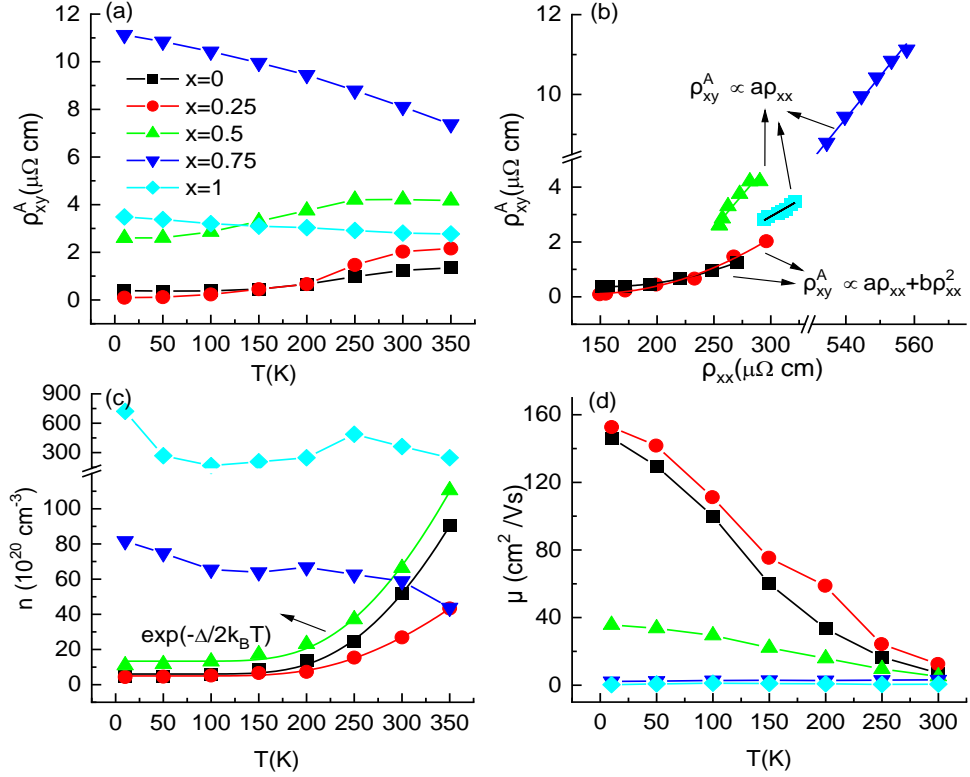

Fig. 4 (a, b) The extracted anomalous Hall resistivity $\rho_{xy}^A$ for $Mn_2CoGa_{1-x}Al_x$ (x=0, 0.25, 0.5, 0.75, 1) in function of temperature and longitudinal resistivity $\rho_{xx}$, respectively. The lines in (b) are fitted curves of the indicated formula. (c, d) The extracted temperature dependences of the carrier concentration $n$ and carrier mobility $\mu$, respectively. The smooth black line in (c) represents the fitted curve of $n = n_0 e^{-\Delta/2k_BT} + const$.

In Fig. 4(b), we also show the relation of $\rho_{xy}^A$ vs $\rho_{xx}$, which can help illustrate the scattering mechanism of AHE. The variation of Ms is disregarded when fitting the curve, since it is exceedingly small within the temperature range of 10-300K (as mentioned before) and has a minimal impact on the final fitted relationship. According to the conventional theory of AHE, $\rho_{xy}^A$ can be scaled with $\rho_{xx}^\alpha$, with $\alpha = 1$ indicating skew scattering and $\alpha = 2$ referring to side jump mechanism or intrinsic Berry phase contribution[30,31]. It is found that for x=0 and 0.25, the $\rho_{xy}^A$ can only be described with a mixed relation of $\rho_{xy}^A = a'\rho_{xx0} + a''\rho_{xxT} + b\rho_{xx}^2$, where the first term refers to the impurity scattering caused by residual resistivity according to the TYJ model [31], the second term refers to the phonon induced resistivity, and the third term indicates the above-mentioned side jump mechanism or intrinsic Berry phase contribution. Since the obtained $a''$ parameter cannot be neglected in our fitting, the side jump contribution cannot be separated from the intrinsic Berry curvature contribution, like ref. [31] do. For x=0.5, 0.75 and 1, it turns out only linear fit ($\rho_{xy}^A = a'\rho_{xx0} + a''\rho_{xxT}$) is reasonable, meaning domination of the extrinsic skew scattering. Therefore, in the high conductive sample (x=0, 0.25), the Hall resistivity can include the intrinsic contribution, which tend to be zero according to the electronic structure calculations, as demonstrated in the subsequent calculation section and ref. [15]. This situation occurs in the dilute magnetic semiconductor, where in the relatively high conductive regime the intrinsic Berry-phase contribution seems to the dominate one in the scattering [32].

Temperature variations of the magnetoresistance (MR = $\frac{\rho(B)-\rho(0)}{\rho(0)}$) for all the samples are also investigated, as seen in Fig. 5. There are several features that should be noted: i) The overall magnitude of MR in x=0 and 0.25 is larger than that in x=0.5-1; ii) There is an apparent positive dip at low temperatures of x=0 -0.5, which will gradually reduce and finally disappear at high temperature; iii) MR at higher fields (usually above the saturation field) will turn from positive to negative with increasing temperature, and in x=0.5, 0.75 and 1, they are completely negative. It is known that in ferromagnets, the MR is affected by both the ordinary cryotron motion and spin-dependent scattering of electrons, where the former contributes positive MR by scaling of $(B/\rho_0)^2$ (according to a two-band model) and the latter normally generates quasi-linear negative MR above the saturation field[33]. Therefore, the large positive MR signal in x=0 and 0.25 should result from the Lorentz effect, owing to the lower $\rho_0$ of them, especially at low temperature, where obvious parabolic relation can be observed. At magnetic field roughly above the saturation point, a downturn or sign reversal of MR takes place due to the rapidly increased contribution of the spin scattering effect. The low-field dip of MR can be attributed to the steep increase of *M* before saturation field, leading to a fast-growing effective field ($B_{eff} = \mu_0(M + H)$). While in x=0.5, 0.75 and 1, the positive contribution from the ordinary MR becomes less significant, thus diminishing the MR magnitude, especially in x=0.75 and 1. It should be mentioned that in Mn$_2$CoAl (x=1), positive MR has also been occasionally observed[2,34], while in other cases only negative MR presented [3,5,35], dependent on the sample condition. The MR evolution trend here reflected the lowering $\rho_0$ or higher carrier mobility $\mu$ in Ga-rich Mn$_2$CoZ alloys. This evokes the possibility that the MR of Mn$_2$CoGa can be higher than Mn$_2$CoAl, when the preparation condition is optimized.

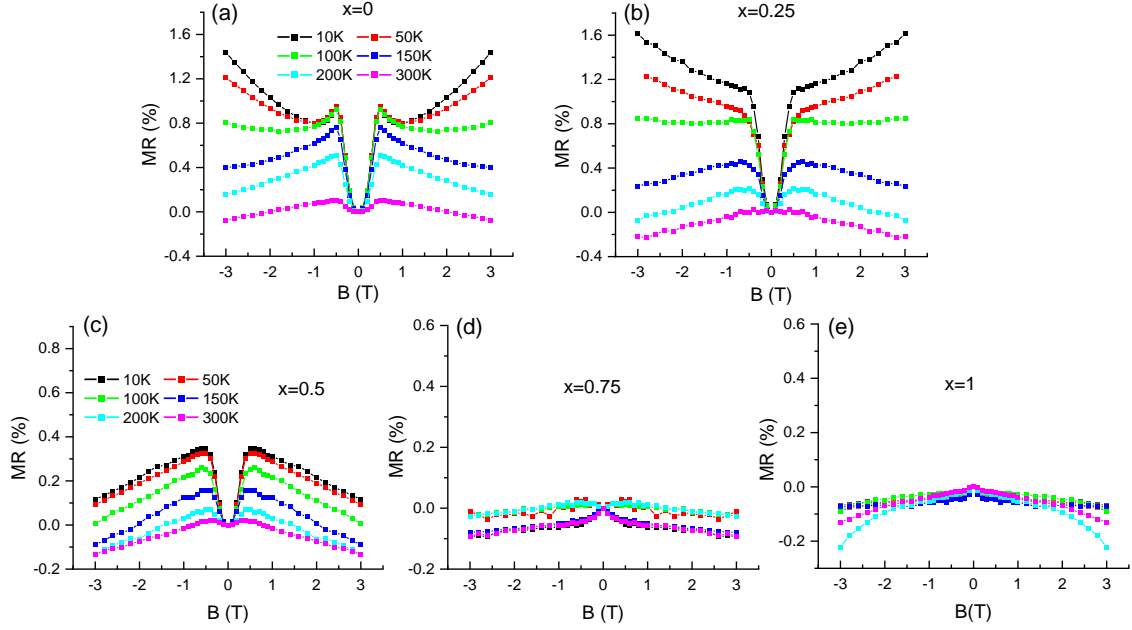

Fig. 5 (a-d) Magnetoresistance for $Mn_2CoGa_{1-x}Al_x$ (x=0, 0.25, 0.5, 0.75, 1) alloys at various temperatures ranging from 10K-300K.

To elucidate the distinct transport properties caused by the mere substitution of the adjacent main-group element, we have examined the electronic structures of $Mn_2CoGa_{1-x}Al_x$ (x=0,0.5,1) alloys using the CPA method (see methods for details). In the fully ordered crystal structure (Fig. 6a), as expected, the substitution has little effect on the density of states (DOS), with the SGS characteristics (near zero DOS value at the Fermi level) preserved. Additionally, we have considered the partial Mn-Co anti-site disorder, which is one of the most common types of defects based on experimental results from $Mn_2CoAl$[6,13] and $Mn_2CoGa$[27,36]. Since Mn occupies two inequivalent sites, A(0,0,0) and B(1/4,1/4,1/4), we have calculated the impact of antisite disorder on $Mn_A$ and Co, as well as $Mn_B$ and Co, and found that the former has a lower energy. Hence, in Fig. 6b, the DOS for the structure with 20% $Mn_A$-Co antisite disorder are presented. It is evident that while the spin down DOS remains unchanged, the spin up DOS increases significantly, destroying the SGS structure. It is worth noting that in the presence of disorder, the DOS for the three alloys are almost identical. Therefore, the difference in transport properties can only be explained by the varying degree of disorder in the Ga substitution. As mentioned in the introduction and also proved in our experiments, spontaneous non-stoichiometry and antisite disorder are more likely happen in $Mn_2CoAl$[6,13], resulting in the lower conductivity and extrinsic scattering mechanism of AHE instead of the intrinsic one. This may be the reason why no vanishing of AHE has ever been observed in $Mn_2CoAl$. For $Mn_2CoGa$, there are also studies showed that it can crystallize to the less ordered $L2_{1b}$ phases, where Mn and Co being partially disordered[27,36]. But it is worth to point out that Ga contained Heusler alloys are generally more ordered than its Al counterpart, like proved by the detailed structural studies on $Co_2FeZ$ (Z=Al, Si, Ga, and Ge)[19], where $Co_2FeAl$ crystallizes in the B2 structure whereas $Co_2FeGa$ crystallizes in the high-order $L2_1$ structure. In addition, the possible non-stoichiometry and phase segregation that cannot be excluded from the XRD can also affect the electronic structure. A more thorough investigation on the microstructure of these alloys is deserved in future to illustrate the properties here, which is out of the scope of this article.

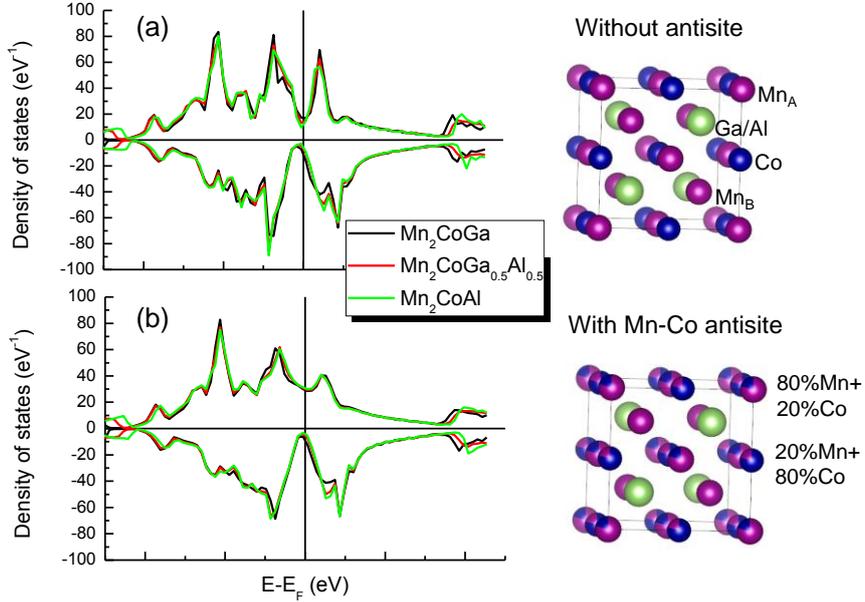

Fig. 6 Spin-resolved density of states for $Mn_2CoGa$, $Mn_2CoGa_{0.5}Al_{0.5}$ and $Mn_2CoAl$ in the ideal $Hg_2CuTi$ structure (a) and in the structure with 20% Mn-Co antisite disorder (b).

## IV. CONCLUSIONS

In this work, by investigating the magnetic and magneto-transport properties of iso-structural $Mn_2CoGa_{1-x}Al_x$ (x=0,0.25,0.5,0.75 and 1) alloys, unusual vanishing of the anomalous Hall effect, accompanying with the growing contribution from the ordinary Hall effect are observed in the high Ga content, indicating the semiconducting like properties in ferromagnetic $Mn_2CoGa$ (or x=0.25) alloy. While in high Al content, despite the magnetization behavior is not altered significantly, the Hall resistivity is instead dominated by the anomalous one, just analogous to that widely reported in $Mn_2CoAl$. Substantially low carrier density ($10^{20}$ cm$^{-3}$) and high carrier mobility (40-150 cm$^2$/Vs) are achieved in $Mn_2CoGa$ (or x=0.25) alloy in the temperature range of 10-200K. These also lead to a large dip in the related magnetoresistance at low fields. The distinct electrical transport behavior of $Mn_2CoGa$ and $Mn_2CoAl$ cannot be illuminated by the electronic structures based on a fully-ordered perfect structure, where more extensive structural studies on these two alloys are deserved in future studies. Our work could provide new insights into the studies of Heusler alloy-based SGS, where the main group element interchange is previously thought to exert no significant effect, and offered a good spin gapless semiconductor candidate with higher carrier mobility that can hopefully promote SGS application in the spin-injected related devices.

## ACKNOWLEDGMENTS

This work is supported by National Natural Science Foundation of China (Grant No. 11604148) and Undergraduate Training Program for Innovation and Entrepreneurship in Nanjing University of Science and Technology.

**REFERENCES**

1. X. Wang. Proposal for a New Class of Materials: Spin Gapless Semiconductors. Phys. Rev. Lett **100**, 156404 (2008).
2. S. Ouardi, G. H. Fecher and C. Felser. realization of spin gapless semiconductors Mn$_2$CoAl.pdf. Phys. Rev. Lett **110**, 100401 (2013).
3. M. E. Jamer, B. A. Assaf, T. Devakul and D. Heiman. Magnetic and transport properties of Mn$_2$CoAl oriented films. Appl. Phys. Lett. **103**, 142403 (2013).
4. I. Galanakis, K. Özdoğan, E. Şaşıoğlu and S. Blügel. Conditions for spin-gapless semiconducting behavior in Mn$_2$CoAl inverse Heusler compound. J. Appl. Phys. **115**, 093908 (2014).
5. G. Z. Xu, Y. Du, X. M. Zhang, H. G. Zhang, E. K. Liu, W. H. Wang and G. H. Wu. Magneto-transport properties of oriented Mn$_2$CoAl films sputtered on thermally oxidized Si substrates. Appl. Phys. Lett. **104**, 242408 (2014).
6. H. Tajiri, L. S. R. Kumara, Y. Sakuraba, Z. Chen, J. Wang, W. Zhou, K. Varun, K. Ueda, S. Yamada, K. Hamaya and K. Hono. Structural insight using anomalous XRD into Mn2CoAl Heusler alloy films grown by magnetron sputtering, IBAS, and MBE techniques. Acta Materialia **235**, 118063 (2022).
7. L. Bainsla, A. I. Mallick, M. M. Raja, A. K. Nigam, B. S. D. C. S. Varaprasad, Y. K. Takahashi, A. Alam, K. G. Suresh and K. Hono. Spin gapless semiconducting behavior in equiatomic quaternary CoFeMnSi Heusler alloy. Phys. Rev. B **91**, 104408 (2015).
8. H. R. Fu, C. Y. You, F. Q. Xin, L. Ma, and N. Tian. Electric-field tuning of magnetism in spin gapless semiconductor (SGS)-like CoFeMnSi thin film. Appl. Phys. Lett. **112**, 262406 (2018).
9. L. Bainsla, A. I. Mallick, A. A. Coelho, A. K. Nigam, B. S. D. C. S. Varaprasad, Y. K. Takahashi, A. Alam, K. G. Suresh and K. Hono. High spin polarization and spin splitting in equiatomic quaternary CoFeCrAl Heusler alloy. J. Magn. Magn. Mater. **394**, 82-86 (2015).
10. P. Kharel, W. Zhang, R. Skomski, S. Valloppilly, Y. Huh, R. Fuglsby, S. Gilbert and D. J. Sellmyer. Magnetism, electron transport and effect of disorder in CoFeCrAl. J. Phys. D: Appl. Phys. **48**, 245002 (2015).
11. L. Bainsla, A. I. Mallick, M. M. Raja, A. A. Coelho, A. K. Nigam, D. D. Johnson, A. Alam and K. G. Suresh. Origin of spin gapless semiconductor behavior in CoFeCrGa: Theory and Experiment. Phys. Rev. B **92**, 045201 (2015).
12. S. Ouardi, G. H. Fecher, C. Felser, J. Kübler, Erratum: realization of spin gapless semiconductors: the Heusler compound Mn2CoAl, Phys. Rev. Lett. **122**, 059901(E) (2019).

13  X. D. Xu, Z. X. Chen, Y. Sakuraba, A. Perumal, K. Masuda, L. S. R. Kumara, H. Tajiri, T. Nakatani, J. Wang, W. Zhou, Y. Miura, T. Ohkubo and K. Hono. Microstructure, magnetic and transport properties of a Mn2CoAl Heusler compound. Acta Materialia **176**, 33-42 (2019).

14  G. Liu, X. Dai, H. Liu, J. Chen, Y. Li, G. Xiao and G. Wu. Mn2CoZ (Z=Al, Ga, In, Si, Ge, Sn, Sb) compounds: Structural, electronic, and magnetic properties. Phys. Rev. B **77,** 014424 (2008).

15  K. Manna, L. Muechler, T.-H. Kao, R. Stinshoff, Y. Zhang, J. Gooth, N. Kumar, G. Kreiner, K. Koepernik, R. Car, J. Kübler, G. H. Fecher, C. Shekhar, Y. Sun and C. Felser. From Colossal to Zero: Controlling the Anomalous Hall Effect in Magnetic Heusler Compounds via Berry Curvature Design. Phys. Rev. X **8**, 041045 (2018).

16  H. Akai. Fast Korringa-Kohn-Rostoker coherent potential approximation and its application to FCC Ni-Fe systems. *J. Phys.: Condens. Matter* **1,** 8045(1989)

17  N. H. Long and H. Akai, First-principles KKR-CPA calculation of interactions between concentration fluctuations. J. Phys.: Condens. Matter **19**, 365232 (2007).

18  H. Akai. http://sham.phys.sci.osaka-u.ac.jp/kkr/.

19  B. Balke, S. Wurmehl, G. H. Fecher, C. Felser, M. C. M. Alves, F. Bernardi and J. Morais. Structural characterization of the $Co_2FeZ$ (Z=Al, Si, Ga, and Ge) Heusler compounds by x-ray diffraction and extended x-ray absorption fine structure spectroscopy. Appl. Phys. Lett. **90**, 172501 (2007).

20  Y. J. Zhang, G. J. Li, E. K. Liu, J. L. Chen, W. H. Wang and G. H. Wu. Ferromagnetic structures in $Mn_2CoGa$ and $Mn_2CoAl$ doped by Co, Cu, V, and Ti. J. Appl. Phys. **113**, 123901 (2013)

21  J. Mooij, Electrical conduction in concentrated disordered transition metal alloys, Phys. Status Solidi A **17**, 521–530 (1973).

22  Y. R. You, G. Z. Xu, J. X. Tang, C. Li, J. Liu, Y. Y. Gong, F. Xu. Spin-glass mediated large exchange bias and coercivity in hexagonal Mn1+xCu1-xGa (x = 0.3-0.5) alloys. J. Phys. D: Appl. Phys. **53**, 265001(2020).

23  Q. Q. Zhang, M. J. Yuan, Z. H. Xia, X. Q. Ma, Z. H. Liu. Structure, magnetism and large anomalous Hall effect of hexagonal MnYSn (Y = Ti, Mn and Fe) ribbons. J. Phys. Chem. Solids **170**, 110944 (2022).

24  J. X. Tang, P. H. Wang, Y. R. You, Y D. Wang, Z. Xu, Z. P. Hou, H. G. Zhang, G. Z. Xu, F. Xu. Abnormal low-field M-type magnetoresistance in hexagonal noncollinear ferromagnetic MnFeGe alloy. Rare Metals **41**, 2680–2687 (2022).

25  H. Pandey, R. C. Budhani. Structural ordering driven anisotropic magnetoresistance, anomalous Hall resistance, and its topological overtones in full-Heusler Co2MnSi thin films. J. Appl. Phys. **113**, 203918 (2013)

26  M. Seredina, I. Gavriko, M. Gorshenkov, S. Taskaev, A. Dyakonov, A. Komissarov, R. Chatterjee, V. Novosad, V. Khovaylo. Magnetic and transport properties of as-prepared Mn2CoGa. J. Magn. Magn. Mater. **470**, 55–58(2019).

27  Rie Y Umetsu, M. Tsujikawa, K. Saito, K. Ono, T. Ishigaki, R. Kainuma and M. Shirai. Atomic ordering, magnetic properties, and electronic structure of Mn2CoGa Heusler alloy. J. Phys.: Condens. Matter. **31**, 065801(2019)

28  A. C. Lawson, A. C. Larson, M. C. Aronson, S. Johnson, Z. Fisk, P. C. Canfield, J. D. Thompson, R. B. Von Dreele, Magnetic and crystallographic order in alpha-manganese. J. Appl. Phys. **76**, 7049 (1994)

29  C. Kittel. Introduction to Solid state physics-8th edition. John Wiley & Sons, Inc. (2005)

30  N. Nagaosa, J. Sinova, S. Onoda, A. H. MacDonald and N. P. Ong. Anomalous Hall effect. Reviews of Modern Physics **82**, 1539-1592 (2010).

31  Y. Tian, L. Ye and X. Jin. Proper scaling of the anomalous Hall effect. Phys. Rev. Lett. **103**, 087206 (2009).

32  S. H. Chun, Y. S. Kim, H. K. Choi, I. T. Jeong, W. O. Lee, K. S. Suh, Y. S. Oh, K. H. Kim, Z. G. Khim, J. C. Woo, and Y. D. Park. Interplay between Carrier and Impurity Concentrations in Annealed $Ga_{1-x}Mn_xAs$: Intrinsic Anomalous Hall Effect, Phys. Rev. Lett. **98**, 026601 (2007).

33  T. R. Mcguire, R. I. Potter. Anisotropic Magnetoresistance in Ferromagnetic 3d Alloys. IEEE Transactions on Magnetics **11**, 1018(1975).

34  K. Kudo, A. Masago, S. Yamada, L. S. R. Kumara, H. Tajiri, Y. Sakuraba, K. Hono and K. Hamaya. Positive linear magnetoresistance effect in disordered $L2_1B$-type Mn2CoAl epitaxial films. Physical Review B **103**, 104427 (2021).

35  P. Chen, C. Gao, G. Chen, K. Mi, M. Liu, P. Zhang and D. Xue. The low-temperature transport properties of Heusler alloy Mn2CoAl. Applied Physics Letters **113**, 122402 (2018).

36  Minakuchi K, Umetsu R Y, Kobayashi K, Ishida K and Kainuma R. Phase equilibria and magnetic properties of Heusler-type ordered phases in the Co-Mn-Ga ternary system J. Alloys Compd. **645**, 577(2015).